\documentclass[aps,amsmath,amssymb,floatfix,11pt]{revtex4-1}
\usepackage{graphicx}
\usepackage{hyperref}
\usepackage{color}

\begin{document}
{\begin{center}
{\bf\large{ Tight-binding models in a quasiperiodic optical lattice}}\\
Anuradha Jagannathan$^{a*}$ and Michel Duneau$^b$ \\
{\it{$^a$ Laboratoire de Physique des Solides, Universit\'e Paris-Sud, 91405 Orsay, France\\
 $^b$ Centre de Physique Th\'eorique, Ecole Polytechnique, CNRS, 91128 Palaiseau, France}} \\
$^*$ Corresponding author email: jagannathan@lps.u-psud.fr \\
\end{center}}

\bigskip
\noindent
{\bf{Introduction}}\\
This paper describes how one can use four standing wave laser fields to realize a two dimensional optical quasicrystal with eight-fold symmetry, closely related to the well-known octagonal  or Ammann-Beenker tiling quasicrystal \cite{beenker}. We describe the structure and its properties, and the effective tight-binding model for atoms in this optical quasicrystal. Such a system, if realized experimentally, should provide valuable insights into the quantum properties of quasicrystals. This would represent a significant progress, since there are very few experimental realizations of simple quasiperiodic structures, whereas there have been many theoretical studies of the physical properties of quasicrystals for tilings in one, two and three dimensions. 

\begin{figure}[!ht]
\centering
\includegraphics[width=120pt]{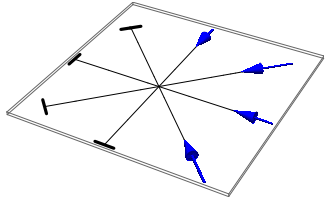}
\caption{Laser and mirror configuration in the $xy$ plane}
\label{laser.fig}
\end{figure}

\bigskip
\noindent
{\bf{Laser intensity field in the $xy$ plane}}\\
We consider a region  in the $xy$ plane where standing waves have been set up using four laser beams, and four mirrors, as shown in the Fig.\ref{laser.fig}. We assume that all the beams are polarized in the direction perpendicular to this plane. Assuming that all the lasers are of equal intensity $I_0$, but with different phase shifts, 
$\phi_n$, the total intensity is given by
$I(\vec{r}) =  I_0 \left[\sum_{n=1}^4 \cos(\vec{k}_n.\vec{r}+ \phi_n)\right]^2 $,
where $\vec{r}=(x,y)$ and the four wave vectors $\vec{k}_n$, directed along angles $n\pi/4$,  have the norm $k=2\pi/\lambda$ where $\lambda$ is the laser wavelength.
An atom subjected to such a standing wave can be preferentially attracted to either the local maxima (''red detuning"), or the minima (''blue detuning") of the laser field (see \cite{bloch} for details). As defined above, $I(\vec{r}) $ is an example of a quasiperiodic function, the theory of which goes back to H. Bohr and A. Besicovitch (\cite{bohr,besi}). Quasiperiodic potentials have been studied before:  for interacting bosonic atoms in one dimension \cite{deiss}, or with 10-fold symmetry \cite{guidoni,sanchez} as well as for colloidal systems, as in \cite{bechinger}. However there has not been, to our knowledge, any experimental realizations of an optical quasicrystal, with information on atomic positions, nor of a corresponding tight-binding model.

\begin{figure}[!ht]
\centering
\includegraphics[width=180pt]{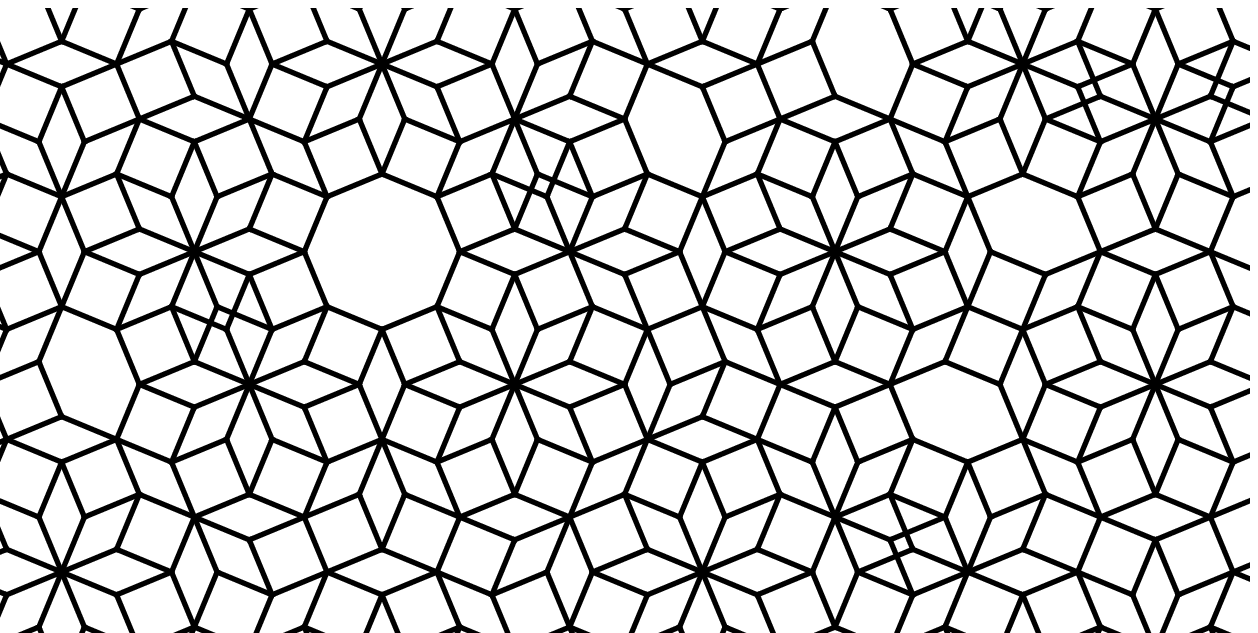} \hskip 2cm
\includegraphics[width=120pt]{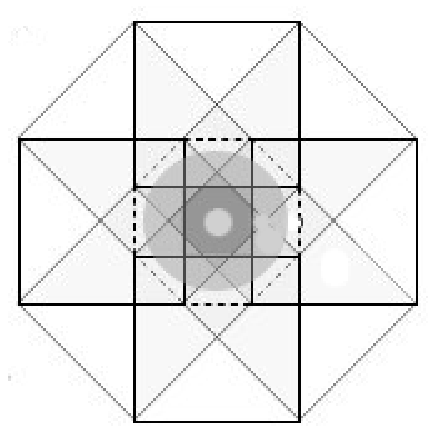}
\caption{(left) a. Optical quasicrystal tiling for $(I_m-I_c)/I_m= 0.17$ showing some of the defect configurations (see text).  (right) b. The octagonal window for the OT is shown, along with three concentric grey circular windows for the OQ for three special values of $I_c$ }
\label{one.fig}
\end{figure}

For strong enough potential, and low enough temperature, atoms will be localized at the local maxima, in the case of a red-detuned lattice. We assume that the occupied sites are those local maxima for which $ I(\vec{r}_j) \geq I_c$ where $I_c$ is a cut-off. Fig.\ref{one.fig}a shows an example of the type of structure obtained for a particular choice of the cut-off. The edge-length  depends only on the laser wavelength $\lambda$ and $\alpha =1+\sqrt{2}$ is the ''silver mean".  As the cut-off $I_c$ increases and approaches the maximum value $I_m=16\vert I_0\vert$, fewer peaks are occupied, and the distance between occupied sites increases.
Fig.\ref{one.fig}a shows that the tiles appearing in the optical system are those found in the standard octagonal tiling (OT), whose eight-fold symmetry and self-similarity under inflation are shared by the optical quasicrystal (OQ).  

\bigskip
\noindent
{\bf{Four dimensional model for the optical quasicrystal}}\\
The wave vectors of the laser beams, $\vec{k}_n$, can be regarded as projections in the $xy$ plane of orthogonal four-dimensional vectors $\vec{K}_n$ where $K=\sqrt{2} k$. Whereas two dimensional lattices do not possess eight-fold rotational symmetry, the four dimensional hypercubic lattice does. The 4D space is, moreover, the direct sum of two orthogonal invariant planes  $P$ and $P'$, having an irrational orientation with respect to the standard basis. One can write
 $\vec{R}=(\vec{r},\vec{r}')$, where  $\vec{r}$ is the projection of a given point in $P$, and $\vec{r}'$ is its projection in $P'$.
If $\vec{K}=(\vec{k},\vec{k}')$ is another vector, the scalar product writes 
$\vec{K}.\vec{R}=\sum K_n R_n=\vec{k}.\vec{r}+\vec{k}'.\vec{r}'$ by orthogonality of $P$ and $P'$. \\

The optical intensity in the $xy$ plane can be obtained from a 4D periodic function :  
$\mathcal{I}(\vec{R})=I_0\left[\sum_{n=1}^4 \cos(\vec{K}_n.\vec{R})\right]^2$ (where the phases are not written as their effect amounts to a global 4D translation). The maxima of this function lie on the vertices of a body centered cubic (BCC) lattice, whose basis vectors, when projected in the $xy$ plane are turned by angles of $3\pi/8$ and $\pi/8$ with respect to the $x$ axis, as can be seen in the tiling of  Fig.\ref{one.fig}a.  
For large $I_c$, the selection rule, $I(\vec{r}_j)\geq I_c$  defines small, approximately spherical regions  around every point of the BCC lattice. These correspond to ''atomic surfaces" in the cut-and-project method, and their projection onto P' defines the selection window, D, a nearly circular disk. The selection window of
the standard octagonal tiling is an octagon W, shown in Fig.\ref{one.fig}b, with a series of smaller (by even powers of $\alpha$) octagons for the inflated tilings. The optical quasicrystal is formed for $I_c$ such that D and W have the same area (see Fig.\ref{one.fig}b)). Further details on this theoretical model can be found in \cite{aj2013}. 
The OT and the OQ thus differ slightly due to their selection windows, giving rise to defects, as can be seen in Fig.1, namely  i)   some missing sites, whence the empty octagons and hexagons, and ii) some close-spaced twin-pairs. Such pairs of sites are conjugates under a phason flip (local atomic jump due to phason modes in the quasicrystal \cite{hen}).  Adding higher harmonics to $I(\vec{r})$ would ensure a better overlap of windows. Alternatively, one could add repulsive interactions between particles to achieve the same result.

\bigskip
\noindent
{\bf{Effective tight-binding model }}\\
At low temperature, atoms occupy the lowest energy state of their potential wells and one can use the basis set of Wannier type localized wavefunctions. In this basis the diagonal matrix elements of the Hamiltonian are $\epsilon_i =\langle i\vert H\vert i\rangle$, and the off-diagonal elements, $t_{ij}=-\langle i\vert H\vert j\rangle$ correspond to the amplitude of tunnelling between sites $i$ and $j$ and depends very strongly on the pair of sites.
The simplest noninteracting model of particles in the OQ is thus described by a hopping Hamiltonian of the form
\begin{equation}
H = -\sum_{\langle i,j\rangle} t_{ij} (a^\dagger_{i} a_{j} + a^\dagger_{j} a_{i}) +\sum_{i=1}^N   \epsilon_i a^\dagger_{i} a_{i},
\end{equation}
where the operator $a_{i} (a^\dag_i)$ annihilates(creates) a particle at site $i$ of the OQ, and sites are labeled $i,j=1,...N$ where $N$ is the total number of lattice sites. In the kinetic (first) term, it is sufficient to consider a small subset of hoppings between nearby sites $i$ and $j$. In the OQ, the two smallest distances are the short diagonal of the rhombus, and the edge. (A shorter distance, between twin-pairs, is infrequent and can moreover be eliminated by taking a smooth cutoff, and turning on a weak repulsive interaction between atoms).
The determination of the typical values of hopping amplitudes for edge and diagonal hops are left for future work. We note that many results are known for the edge hopping Hamiltonian on the octagonal tiling: for the spectrum and wavefunctions \cite{sire1,sire2,zhong}, for local densities of states and RKKY interactions \cite{localdos}, for the statistics of the energy levels (reviewed in \cite{philmag}), for quantum dynamics \cite{ben2}, effect of Hubbard interaction and the Heisenberg model \cite{jagsch,wess,jagprl}.

\bigskip
\noindent
{\bf{Summary and discussion}}
We have discussed the structure and tight-binding model of an 8-fold optical quasicrystal and related it to the well-known octagonal tiling.  It is in principle easier to realize experimentally, as well as being simpler conceptually, than 10-fold systems generated using five laser beams, and so, provides an ideal system in which to study the classical and quantum physics of quasiperiodic structures. 

\bigskip
\noindent
{\bf {Acknowledgments}}
We wish to thank Christoph Weitenberg (LKB, Paris), Bess Fang (Institute of Optics, Palaiseau) and Monika Aidelsburger (LMU, M\"unchen), as well as J.-M. Luck and J-.F. Sadoc for useful discussions.

\end{document}